# Skin Effect of Nonlinear Optical Responses in Antiferromagnets

Hang Zhou,[1,2] Rui-Chun Xiao,[1,3*] Shu-Hui Zhang,[4] Wei Gan,[1] Hui Han,[1,3] Hong-Miao Zhao,[1] Wenjian Lu,[2]
Changjin Zhang,[1,5] Yuping Sun,[5,2,6] Hui Li,[1,3†] and Ding-Fu Shao[2,‡]

[1] *Institute of Physical Science and Information Technology, Anhui University, Hefei 230601, China*

[2] *Key Laboratory of Materials Physics, Institute of Solid State Physics, Hefei Institutes of Physical Science,*
*Chinese Academy of Sciences, Hefei 230031, China*

[3] *Anhui Provincial Key Laboratory of Magnetic Functional Materials and Devices, Anhui University, Hefei 230601, China*

[4] *College of Mathematics and Physics, Beijing University of Chemical Technology, Beijing 100029, China*

[5] *Anhui Province Key Laboratory of Low-Energy Quantum Materials and Devices, High Magnetic Field Laboratory,*
*Hefei Institutes of Physical Science, Chinese Academy of Sciences, Hefei 230031, China*

[6] *Collaborative Innovation Center of Microstructures, Nanjing University, Nanjing 210093, China*

[*]xiaoruichun@ahu.edu.cn; [†]huili@ahu.edu.cn; [‡]dfshao@issp.ac.cn

Nonlinear optics plays important roles in the research of fundamental physics and the applications of high-performance optoelectronic devices. The bulk nonlinear optical responses arise from the uniform light absorption in noncentrosymmetric crystals, and hence are usually considered to be the collective phenomena of all atoms. Here we show, in contrast to this common expectation, the nonlinear optical responses in antiferromagnets can be selectively accumulated near the surfaces, representing a skin effect. This is because the inversion symmetry, despite being broken globally by magnetism, is barely violated locally deeply inside these antiferromagnets. Using A-type layered antiferromagnets as the representatives, we predict that the spatial-dependent nonlinear optical responses, such as bulk photovoltaic effect (BPVE) and second harmonic generation (SHG), are notable in the top- and bottom-most layers and decay rapidly when moving away from the surfaces. Such a phenomenon is strongly associated with the antiferromagnetism and exists in a broad range of antiferromagnets composed of centrosymmetric sublattices, offering promising device applications using these antiferromagnets. Our work uncovers a previously overlooked property of nonlinear optical responses and opens new opportunities for high-performance antiferromagnetic optospintronics.

The interplay between light and matter has given rise to intriguing physical phenomena that significantly advance sciences and technology [1,2]. Particular interests have been attracted by the nonlinear optical effects [3,4] due to the fascinating excitation processes and the associated applications. Typical examples are bulk photovoltaic effect (BPVE) and second harmonic generation (SHG), which involve the geometric phase of the Bloch wavefunctions and are proven to be powerful probes of symmetry and topology [2,4-7] related properties difficult to be detected using other experimental probes. These nonlinear optical effects require the broken inversion symmetry ($\hat{P}$), and hence are long believed to emerge only in materials with noncentrosymmetric crystal structures.

Recently, it was found that nonlinear optical effects such as BPVE and SHG can also emerge in materials with centrosymmetric crystal structures, provided their $\hat{P}$ symmetry is broken by the magnetic alignments. It has been confirmed in antiferromagnets where the magnetizations are compensated by the combined $\hat{P}\hat{T}$ symmetry, even though the $\hat{P}$ and the time reversal $\hat{T}$ symmetry operations are independently broken [8-12]. The nonlinear optical effects in such antiferromagnets show distinct behavior compared to that in noncentrosymmetric ($\hat{T}$ preserved) crystals. First, the nonlinear optical effects are time-reversal-odd ($\hat{T}$-odd) in such antiferromagnets. Therefore, they show notable response against the switching of antiferromagnets, and hence can be used to detect the antiferromagnetic Néel vectors [13-16] and domain states [7,17,18], which are essential for efficient information read-out in antiferromagnetic spintronics [10,19]. Moreover, the nonlinear optical effects in these systems have different origins. In conventional noncentrosymmetric ($\hat{T}$ preserved) crystals, the BPVE and SHG involve the circular injection currents and the linear shift currents, while in $\hat{P}\hat{T}$ symmetric antiferromagnets are associated with the linear injection currents and the circular shift currents [10-12,20]. Despite various theoretical [8,11,12,14,21,22] and experimental [9,13,23-26] progress, the investigations of nonlinear optical effects in antiferromagnets are still in a rudimentary stage. It would be interesting from the fundamental point of view and desirable for optoelectronics/optospintronic[27] applications to explore more unique features of the nonlinear optical effects in antiferromagnets.

Conventionally, the properties of materials are understood based on the analyses of their global symmetry. On the other hand, recently, it was found that there are various physical phenomena hidden within the macroscopic properties [28,29], which cannot be derived simply from the global symmetry analyses but require the investigations of the site-dependent symmetry and properties. For example, it was found that antiferromagnets can host sublattice-dependent transport spin polarizations, resulting in Néel spin currents hidden in a globally charge current [29,30], even the global spin polarization might be vanishing. This can be exploited to drive spin-transfer torques and sizable tunneling magnetoresistance (TMR) effects in antiferromagnetic tunnel junctions [31]. Similarly, one may expect the interesting but previously unknown features of other macroscopic properties, such as nonlinear optical effects, revealed in antiferromagnets.

In this work, we demonstrate the previously overlooked skin effect of the nonlinear optical responses in antiferromagnets. Based on symmetry analyses and theoretical calculations, we show that despite the global $\hat{P}$ breaking by magnetism, the second-order nonlinear optical responses such as BPVE and SHG in antiferromagnets are selectively contributed by the atoms close to the surfaces, provided the local $\hat{P}$ symmetry deeply inside these antiferromagnets are barely violated. Using the $\hat{P}\hat{T}$ symmetric A-type layered antiferromagnet CrI$_3$ as the representative, we show the spatial dependent nonlinear optical coefficients, such as the BPVE coefficient and the SHG coefficient, are notable in the top- and bottom-most layers and decay rapidly when moving away from the surfaces. We argue such a skin effect exists not only in the $\hat{P}\hat{T}$ symmetric layered antiferromagnets, but also in a broad range of antiferromagnets composed of centrosymmetric sublattices.

We begin with the analyses on BPVE, a typical nonlinear optical effect that converts the photons to electric currents [10,32-34]. The photocurrent along $a$-direction generated by BPVE is $J_a = \sigma_{bc}^a E_b E_c$, where $\sigma_{bc}^a$ is the BPVE coefficient, $E_b$ and $E_c$ are the electric fields of the light along $b$ and $c$ directions, respectively. $J_a$ originates from the interband transitions, and can host the two components, i.e. the injection and shift photocurrents, depending on the types of light driving the BPVE [10-12,20]. For simplification, here we focus on the injection current, which is the basic type of the photocurrent in $\hat{P}\hat{T}$ symmetric antiferromagnets generated by a linear polarized light [2,10-12,22]. It originates from the change of carrier group velocity when they are excited from the initial to the final states. The linear injection current coefficient can be calculated as [2,10-12,20,22]

$$\sigma_{bc}^a(0,\omega,-\omega) = -\frac{\pi e^3 \tau}{2\hbar^2} \int \frac{d^3 \mathbf{k}}{(2\pi)^3} \sum_{n,m} f_{nm} \Delta_{nm}^a g_{bc}^{mn} \delta(\omega_{mn} - \omega), \quad (1)$$

where $f_{nm}$ is the Fermi occupation difference of the $n$-th and $m$-th bands, $\omega$ is the frequency of the light, $\omega_{mn}$ is the energy difference of the $m$-th and the $n$-th band, $\Delta_{nm}^a = v_{nn}^a - v_{mm}^a$ is the band velocity difference, and $\tau$ is the relaxation time. $g_{ba}^{mn} = \{r_{nm}^b, r_{mn}^a\}$ is the quantum metric, and $r_{mn}^a$ is inter-band Berry connection. $g_{ba}^{mn}$ ($\Delta_{nm}^a$) is even (odd) with respect to both $\hat{P}$ symmetry and $\hat{T}$ symmetry. Therefore, $\sigma_{bc}^a$ vanishes in centrosymmetric nonmagnets. However, if a centrosymmetric crystal host collinear antiferromagnetism which breaks both $\hat{P}$ and $\hat{T}$ symmetry, the $\sigma_{bc}^a$ can be finite, despite the combined $\hat{P}\hat{T}$ symmetry being preserved [2,10-12,22].

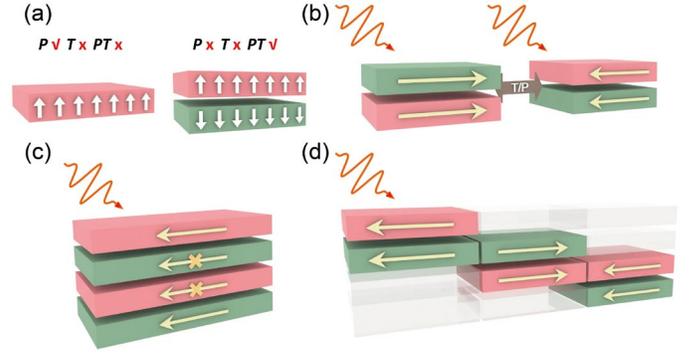

**Fig. 1:** (a) *Left*: Schematic of a 2D centrosymmetric monolayer ferromagnet. *Right*: Schematic of a bilayer A-type antiferromagnet composed of two antiparallel-aligned ferromagnetic monolayers shown in left. (b) Schematic of the BPVE in a bilayer A-type antiferromagnet, where the layer-resolved photocurrent (denoted by the yellow arrows) is identical in each layer and reversable by the Néel vector switching. (c, d) Schematic of the skin effect of BPVE in a four-layer A-type antiferromagnet, where the photocurrent is symmetrically accumulated in the top and bottom layers (c). This can be intuitively understood by considering that the four-layer antiferromagnet is constructed by three overlapping bilayers, where the BPVE in the inner the overlapping layers are compensated (d).

The recently discovered two-dimensional (2D) A-type van der Waals antiferromagnets, such as CrX$_3$ (X = Cl, Br, and I) [35-38], CrSBr [24], and MnBi$_2$Te$_4$ [39-42] are typical $\hat{P}\hat{T}$ symmetric antiferromagnets. The basic building blocks of these antiferromagnets are ferromagnetic monolayers, which host the $\hat{P}$ symmetry but the $\hat{T}$ symmetry is broken by the ferromagnetism (Fig. 1(a)). The antiferromagnetic interlayer exchange coupling results in the alternating stacking of the antiparallel-aligned monolayers, where the $\hat{P}$ symmetry is broken and the $\hat{P}\hat{T}$ symmetry is enforced in the even-layer stacking. This allows the BPVE (linear injection current) in such layered antiferromagnets [20,22,23,43,44].

Figure 1(a) shows such an $\hat{P}\hat{T}$ symmetric antiferromagnetic bilayer. The spatial distribution of the photocurrent generated by BPVE in such a bilayer can be decomposed to the layer-resolved injection current coefficients $\sigma_{bc,i}^a$

$$\sigma_{bc,i}^a = \sum_j \sigma_{bc,ij}^a = -\frac{\pi e^3 \tau}{2\hbar^2} \int \frac{d^3 \mathbf{k}}{(2\pi)^3} \sum_{n,m,j} f_{nm} \Delta_{nm,ij}^a g_{bc}^{mn} \delta(\omega_{mn} - \omega), \quad (2)$$

where $i$ and $j$ are the layer indexes, $\sigma_{bc,ij}^a$ is the injection current coefficient contributed by the transition from the $j$-th layer to the $i$-th layer, $\Delta^a_{nm,ij} = \langle n|P_i\hat{v}^a P_j|n\rangle - \langle m|P_i\hat{v}^a P_j|m\rangle$, and $P_i$ is the projection operator [45]. Obviously, the photocurrents within the bottom layer ($i = 1$) and top layer ($i = 2$) is equivalent, which is enforced by the symmetry transformation $\hat{P}\hat{T}\sigma^a_{bc,1} = \sigma^a_{bc,2}$ in such an antiferromagnetic bilayer [46]. Switching the Néel vector is corresponding to the application of a $\hat{T}$ or $\hat{P}$ symmetry operation on this bilayer, which reverses the photocurrents as $\hat{P}\sigma^a_{bc,1} = -\sigma^a_{bc,2}$ or $\hat{T}\sigma^a_{bc,i} = -\sigma^a_{bc,i}$ (Fig. 1(b)) [46].

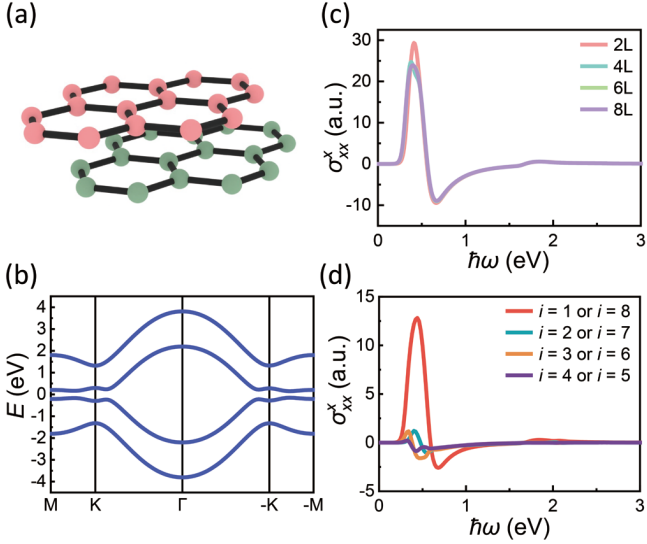

**Fig. 2:** (a) The structure of the A-type antiferromagnet used to build the toy model, which is composed of antiparallel-aligned ferromagnetic honeycomb monolayers. Here only two adjacent layers are shown for simplification. (b) The band structure of the toy model with a thickness of two monolayers. (c) The total BPVE coefficient $\sigma^x_{xx}$ of the toy model with the thicknesses of 2, 4, 6, and 8 layers. (d) The layer-resolved BPVE coefficient $\sigma^x_{xx,i}$ for the $i$-th layer of the toy model with the thicknesses of 8 layers.

One may expect that the photocurrent would be stronger and remain uniform in a thicker antiferromagnetic slab, since the linear optical absorption is identical for each layer [45]. However, in contrast to this intuitive conjecture, we find that the photocurrent is accumulated in the layers near the surfaces (Fig. 1(c)), despite the global $\hat{P}$ breaking. In order to demonstrate this unexpected phenomenon, here we take a four-layer A-type antiferromagnet as an example, and naively consider that it is constructed by three overlapping bilayers, as shown in Fig. 1(d). Obviously, the Néel vector and hence the photocurrent of the central bilayer are opposite to these in the top and bottom bilayers. As a result, the photocurrent is compensated within the overlapping layers and is sustained only within the top and bottom layers. This is analogous the skin effect of an AC current non-uniformly distributed over the surface of the conductor.

Such a skin effect of the photocurrent can be more precisely revealed by performing a general analysis on $\sigma^a_{bc,ij}$ in Eq. (2), which is contributed by the intralayer transition when $j = i$ and the interlayer transition from the $m$-th layer above ($+$) or below ($-$) the layer $i$ when $j = i \pm m_\pm$. Since the ferromagnetic monolayers are centrosymmetric, $\sigma^a_{bc,i}$ must be vanishing for a bulk with infinite layers ($m_\pm = \infty$), because $\sigma^a_{bc,i(i-m)} = -\sigma^a_{bc,i(i+m)}$ enforced by $\hat{P}$ symmetry, where $m$ is an aribitrary integer. On the other hand, $\sigma^a_{bc,i}$ should be nonvanishing in a $n$-layer thin slab where $n$ is even, since $m_\pm$ are finite and $m_+ \neq m_-$. However, due to the weak interlayer coupling in the van der Waals layered materials, there might be a critical number $m_0$ to ensure that only the $\sigma_{i \pm m_\pm}$ is nonnegligible for $m_\pm \leq m_0$. For example, if we assume that the interlayer contribution is only from the nearest neighbor layers ($m_0 = 1$), we have

$$\sigma^a_{bc,i} \approx \sigma^a_{bc,i(i-1)} + \sigma^a_{bc,ii} + \sigma^a_{bc,i(i+1)}. \quad (3)$$

For the layers far away from the surface with bulk-like structures, the relation $\sigma^a_{bc,i(i-m)} = -\sigma^a_{bc,i(i+m)}$ may be barely violated and thus $\sigma^a_{bc,i} \approx 0$. In another word, although the $\hat{P}$ symmetry is broken globally, it may be roughly preserved locally to prevent the photocurrent in the layers far away from the surface. On the contrary, the $\sigma^a_{bc,i}$ must be finite for the surface layers, since there is only one adjacent layer next to them and hence $\sigma^a_{bc,i(i+1)} = 0$ and $\sigma^a_{bc,i(i-1)} = 0$ for the top ($i = n$) and bottom ($i = 1$) surface layers, respectively. The skin effects are expected for other second-order responses of the 2D A-type antiferromagnets driven by a linear polarized light, such as the SHG [9,14,47]. The skin effect of the shift current and SHG driven by the circular light can be also derived by similar analyses.

In order to numerically illustrate such nonlinear optical skin effects, we first consider a toy model of an $\hat{P}\hat{T}$ symmetric layered A-type antiferromagnet composed of ferromagnetic monolayers (indexed by $i$) with the honeycomb lattice (Fig. 2(a)). Each monolayer is described by the Kane-Mele Hamiltonian [48-50]

$$H_i = t \sum_{p,q,\alpha} c^\dagger_{p\alpha} c_{q\alpha} + \lambda_{M,i} \sum_i c^\dagger_p s^z c_p$$
$$+ i\lambda_{SO} \sum_{\langle\langle p,q\rangle\rangle} \nu_{pq} c^\dagger_p s^z c_q + \lambda_R \sum_{\langle\langle p,q\rangle\rangle} \mu_{pq} c^\dagger_p (\mathbf{s} \times \mathbf{d}_{pq})_z c_q, \quad (4)$$

where $p$ and $q$ are the site indexes, $\alpha$ is the spin index, $\mathbf{s}$ is the Pauli matrix, $\mathbf{d}_{pq}$ is the vector pointing from the $p$-th site toward the $q$-th site, $t$ is the nearest hopping parameter, $\lambda_{M,i} = -\lambda_{M,i+1}$ is the parameter of Zeeman splitting of the $i$-th layer, $\lambda_{SO}$ is the spin orbit coupling (SOC) parameter associated with the spin-dependent hopping, $\lambda_R$ is the intrinsic Rashba SOC parameter preserved in $\hat{P}$ symmetry, $\nu_{pq} = (\mathbf{d}_p \times \mathbf{d}_q)/|\mathbf{d}_p \times \mathbf{d}_q| = \pm 1$, and $\mu_{pq} = \pm 1$ for the $A(B)$ site. The Hamiltonian of a $n$-layer antiferromagnetic slab with the rhombohedral stacking (Fig. 2(a)) is then

$$H = \begin{bmatrix} H_{11} & H_{12} & 0 & 0 & 0 \\ H_{21} & H_{22} & H_{23} & 0 & 0 \\ 0 & H_{32} & H_{33} & H_{34} & \vdots \\ 0 & 0 & H_{43} & \ddots & H_{(n-1)n} \\ 0 & 0 & \cdots & H_{n(n-1)} & H_{nn} \end{bmatrix}, \quad (5)$$

where the diagonal $H_{ii}$ is the Hamiltonian $H_i$ in Eq. (4), and the off-diagonals denote the interlayer interaction. Due to the presence of van der Waals gap, it is safe to assume that the nearest neighbor layer interaction is dominant, hence Eq. (5) is triangular block diagonal.

We take typical parameters $t = 1$ eV, $\lambda_{M,i} = 0.8$ eV, $\lambda_{SO} = 0.1$ eV and $\lambda_R = 0.2$ eV within the monolayer, and consider the hopping parameters $t_N = 0.2$ eV and $t_{NN} = 0.01$ eV between the nearest and next nearest magnetic atoms of the two adjacent layers. For an even-layer slab, this Hamiltonian results in a spin-degenerate band structure enforced by $\hat{P}\hat{T}$ symmetry, with a band gap around Fermi level ($E_F$) opened by SOC (Fig. 2(b)). There are only three nonvanishing in-plane coefficients $\sigma_{xx}^x = -\sigma_{yy}^x = -\sigma_{xy}^y$ for the injection current, due to the existences of a three-fold rotation along $z$-axis ($\hat{C}_{3z}$) and a two-fold rotation around $x$-axis ($\hat{C}_{2x}$). A peak of $\sigma_{xx}^x$ along $\hbar\omega = 0.42$ eV is found (Fig. 2(c)), corresponding to the excitation from the valence band maximum (VBM) to the conduction band minimum (CBM). Surprisingly, we find the $\sigma_{xx}^x$ changes slightly with the variation of thickness.

In order to understand the nature of BPVE in such an antiferromagnetic slab, we calculate the layer-resolved $\sigma_{xx}^x$. We find $\sigma_{xx,i}^x = \sigma_{xx,n-i+1}^x$, as presented in Fig. 2(d) using an 8-layer slab as a representative. Moreover, $\sigma_{xx,i}^x$ slightly changes for different $n$, and $\sigma_{xx,1}^x = \sigma_{xx,8}^x \gg \sigma_{xx,i}^x$ ($1 < i < 8$) (Fig. 2 (d)), which is the typical feature of a skin effect. We also calculate the SHG coefficient $\chi_{xx}^x (= -\chi_{yy}^x = -\chi_{xy}^y)$ and find similar skin effect [45].

We then calculate such a skin effect of nonlinear optical responses in a realistic material CrI$_3$, a 2D A-type layered antiferromagnet discovered recently [35,51]. CrI$_3$ is a van der Waals magnetic insulator where the Cr magnetic moments are ferromagnetic in each monolayer with an easy axis along the [001] direction ($z$-axis) (Fig. 3(a)). The interlayer exchange interaction of CrI$_3$ can be either ferromagnetic or antiferromagnetic, depending on the stacking modes of the interlayers [52-55]. Although CrI$_3$ monolayer has the $\hat{P}$ and $\hat{C}_{3z}$ symmetry, the antiferromagnetic stacking breaks these symmetries in the even-layer slab (Fig. 3(a)). The resultant magnetic space group C2/m' hosts the $\hat{P}\hat{T}$ symmetry and supports three independent in-plane BPVE coefficients $\sigma_{yx}^x$, $\sigma_{xx}^y$ and $\sigma_{yy}^y$ [14,47].

We calculate the coefficients of BPVE and SHG in a slab of antiferromagnetic CrI$_3$ based on the tight-binding models derived from the first-principles density functional theory calculations [45]. Figures 3(b-d) shows the calculated $\sigma_{yx}^x$ for a 10-layer CrI$_3$. We find the layer-resolved $\sigma_{yx,i}^x$ are perfectly symmetric as $\sigma_{yx,i}^x = \sigma_{yx,n+1-i}^x$ ($n = 10$). However, unlike the case in the effective model where the BPVE is dominant only for the surface layers indexed by $i = 1$ and $i = n$, we find the layers indexed by $i = 2$ and $i = n - 1$ in CrI$_3$ also contribute notably to the BPVE (Fig. 3(c)). This may be due to the longer and stronger interlayer hopping distance in realistic materials, leading to a strong violation of $\hat{P}$ symmetry in these layers. On the other hand, the $\sigma_{yx,i}^x$ for the layers deeply inside the slab (e.g., $i = 5, 6$) are much smaller (Fig. 3(c)). Figure 3(d) shows the $\sigma_{yx,i}^x$ that contribute to the most pronounced peak of $\sigma_{yx}^x$ at $\hbar\omega_0 = 1.44$ eV. We find $\sigma_{yx,i}^x(\omega_0)$ is significant for the two outer layers, and then decays rapidly when moving away from the surfaces, which is the typical feature of a skin effect. Such a skin effect is sustained for the $\sigma_{yx}^x$ at other major peaks, and is further quantified by calculating the average $\sigma_{yx,i}^x$, as shown in Fig. S6 in supplemental material [45]. We find other nonlinear optical coefficients $\sigma_{xx}^y$ and $\sigma_{yy}^y$ also host such a skin effect. This behavior also emerges for SHG coefficients [45].

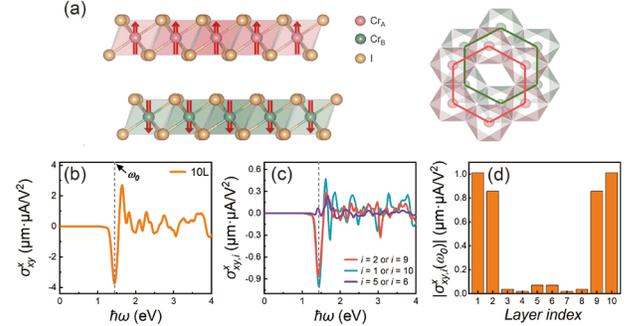

**Fig. 3:** (a) The side view (left) and top view (right) of two adjacent layers of antiferromagnetic CrI$_3$. The total (b) and the layer-resolved (c) BPVE coefficient $\sigma_{xy}^x$ for a 10-layer CrI$_3$. (d) The layer-resolved $\sigma_{xy}^x$ at $\hbar\omega_0 = 1.44$ eV.

Due to the skin effect, the global nonlinear optical coefficients show slight changes when the thickness of the even-layer CrI$_3$ slab is variant. Therefore, the proposed skin effect can be confirmed by detecting the global nonlinear optical responses of A-type layered antiferromagnets with different thicknesses (Fig. 4(a)). In addition, it is possible to make an optoelectronic/optospintronic device as schematically shown in Fig. 4(b), where only the surface layer on one side of a A-type antiferromagnetic slab has contacts with the electrodes. The photocurrent of this surface layer can thus be selectively detected, which would be slightly smaller than one half of the global photocurrent (Fig. 4).

Although we have only calculated the coefficients of BPVE and SHG driven by the linear polarized light, those induced by circular polarized light [2,8,11,12,20] could also host the skin effects [45]. Moreover, the skin effect should also exist in the light-induced spin photocurrent via bulk spin photovoltaic effect [10,21,43,44] and the nonlinear Hall effect [39,40] as well.

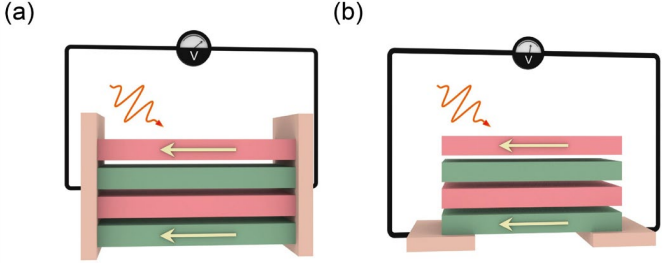

**Fig. 4:** (a) Schematic of an optoelectronic/optospintronic device based on a 2D A-type antiferromagnetic slab, where the electrodes have contacts with all the layers to collect the total photocurrents. (b) Schematic of an optoelectronic/optospintronic device based on a 2D A-type antiferromagnetic slab, where the electrodes have contacts with only the bottom layer to collect the photocurrent of the bottom layer.

Besides, there could be also the *hidden* skin effect associated with the vanishing global nonlinear optical responses, due to the layer-resolved responses for the top and bottom layers being staggered. For example, the layer-resolved photocurrent may be spin-polarized due to the $\hat{T}$ breaking locally, and the resultant layer-resolved spin photocurrents, i.e. the light-induced Néel spin currents, are staggered by $\hat{P}\hat{T}$ symmetry [46]. This prevents a global spin photocurrent, but allows a hidden skin effect of the spin photocurrent, where the light-induced staggered Néel spin currents are accumulated at the two surfaces. Similarly, a hidden skin effect of the charge photocurrent may emerge in the direction the global spin photocurrent is forbidden [10,11,43,45,46]. Such a hidden skin effect can be detected by measuring the responses selectively on a single surface, using the devices schematically shown in Fig. 4(b).

We argue the skin effect and the hidden skin effect of these nonlinear optical responses exist not only in the proposed 2D layered antiferromagnets, but also in a broad range of antiferromagnets, even for these do not host the A-type stacking, the hexagonal structures, or the van der Waals gaps. The symmetry criteria about the skin effect of the nonlinear optical responses in antiferromagnets have been concluded in Supplemental Material [45].

The skin effect of these nonlinear optical responses is promising for optoelectronic/optospintronic applications. First, it makes the nonlinear optical responses nearly thickness-independent, thus allows the observing of these responses in a not-too-thin sample, which avoids the difficulties in experiments to make a high-quality sample down to the 2D limit. Second, such a skin effect would largely simplify the structures of optoelectronic/optospintronic devices, since the electrodes only need to have contacts with one surface layer instead of all the layers (Fig. 4). Third, the skin effect and the hidden skin effect allow exploiting the nonlinear optical responses on upper and lower surfaces independently. One can thus design optospintronic with high density based on this property.

In conclusion, we demonstrate that antiferromagnets can host a previously overlooked skin effect of the nonlinear optical responses, such as BPVE and SHG, provided the local $\hat{P}$ symmetry deeply inside these antiferromagnets is barely violated. Using the $\hat{P}\hat{T}$ symmetric A-type layered antiferromagnets as the representative, we show the spatial dependent nonlinear optical coefficients, such as the BPVE coefficient and SHG coefficient, are notable in the top- and bottom-most layers and decay rapidly when moving away from the surfaces. We argue there could be also hidden skin effects of some nonlinear optical responses, despite they are globally vanishing. Such the skin effect and the hidden skin effect are strongly associated with the antiferromagnetism and exist in a broad range of antiferromagnets composed of centrosymmetric sublattices, offering new opportunities of high-performance optoelectronics and optospintronics using these antiferromagnets. We hope therefore that our predictions will stimulate experimental investigations of the skin effect of nonlinear optical responses and the associated device applications.

**Acknowledgments.** We appreciate the helpful discussion with Xing-Qiang Shi, Naizhou Wang, Haowei Chen, and Yang Gao. We thank the support from the National Key R&D Program of China (Grant No. 2021YFA1600200), the National Natural Science Foundation of China (Grants Nos. 12204009, 12241405, 12274411, 52250418, 12274412, 12204003 and 12474100), the Basic Research Program of the Chinese Academy of Sciences Based on Major Scientific Infrastructures (Grant No. JZHKYPT-2021-08), the CAS Project for Young Scientists in Basic Research (Grant No. YSBR-084), and Natural Science Foundation of Anhui Province (Grant No. 2208085QA08). Computations were performed at Hefei Advanced Computing Center and the High-performance Computing Platform of Anhui University.